\documentclass[twocolumn,twoside]{revtex4}
\usepackage{graphicx}
\usepackage{fancyhdr}
\usepackage{amsmath}
\PassOptionsToPackage{breaklinks}{hyperref}
\usepackage{hyperref}
\hypersetup{	colorlinks=true,
			linkcolor=blue,
			citecolor=blue,
			urlcolor=blue,
		}	
\usepackage{url}
\usepackage[hyphenbreaks]{breakurl}

\begin{document}

\title{Rare Kaon Decays}

%

\author{Chieh Lin}
\affiliation{University of Chicago, Chicago, Illinois, USA, 60637}

\begin{abstract}
The experimental status of the $K \to \pi \nu \overline{\nu}$ search is presented. The $K \to \pi \nu \overline{\nu}$ decay is sensitive to New Physics because it is theoretically pristine and highly suppressed. The $K_L^0 \to \pi^0 \nu \overline{\nu}$ search is performed by the KOTO experiment and a branching fraction limit of $\mathcal{B}(K_L^0 \to \pi^0 \nu \overline{\nu})$ $<$ 3.0 $\times$ 10$^{-9}$ (90\% confidence level) was set. This limit is $\mathcal{O}(100)$ times larger than the Standard Model prediction. The $K^+ \to \pi^+ \nu \overline{\nu}$ search is performed by the NA62 experiment and $\mathcal{B}(K^+ \to \pi^+ \nu \overline{\nu})$ $=$ (10.6 $^{+4.0}_{-3.5}$ $|_{\text{stat.}}$) $\pm$ 0.9$_{\text{syst.}}$) $\times$ 10$^{-11}$ (68\% confidence level) was set. This shows an agreement with the Standard Model. Both measurements can also be used to search for the dark particle $X$ via the $K \to \pi X$ decay. A projection of the $K \to \pi \nu \overline{\nu}$ search in the future is also given. 
\end{abstract}

\maketitle

\thispagestyle{fancy}


%
%


\section{Introduction}
\begin{figure}[h]
\centering
\includegraphics[width=80mm]{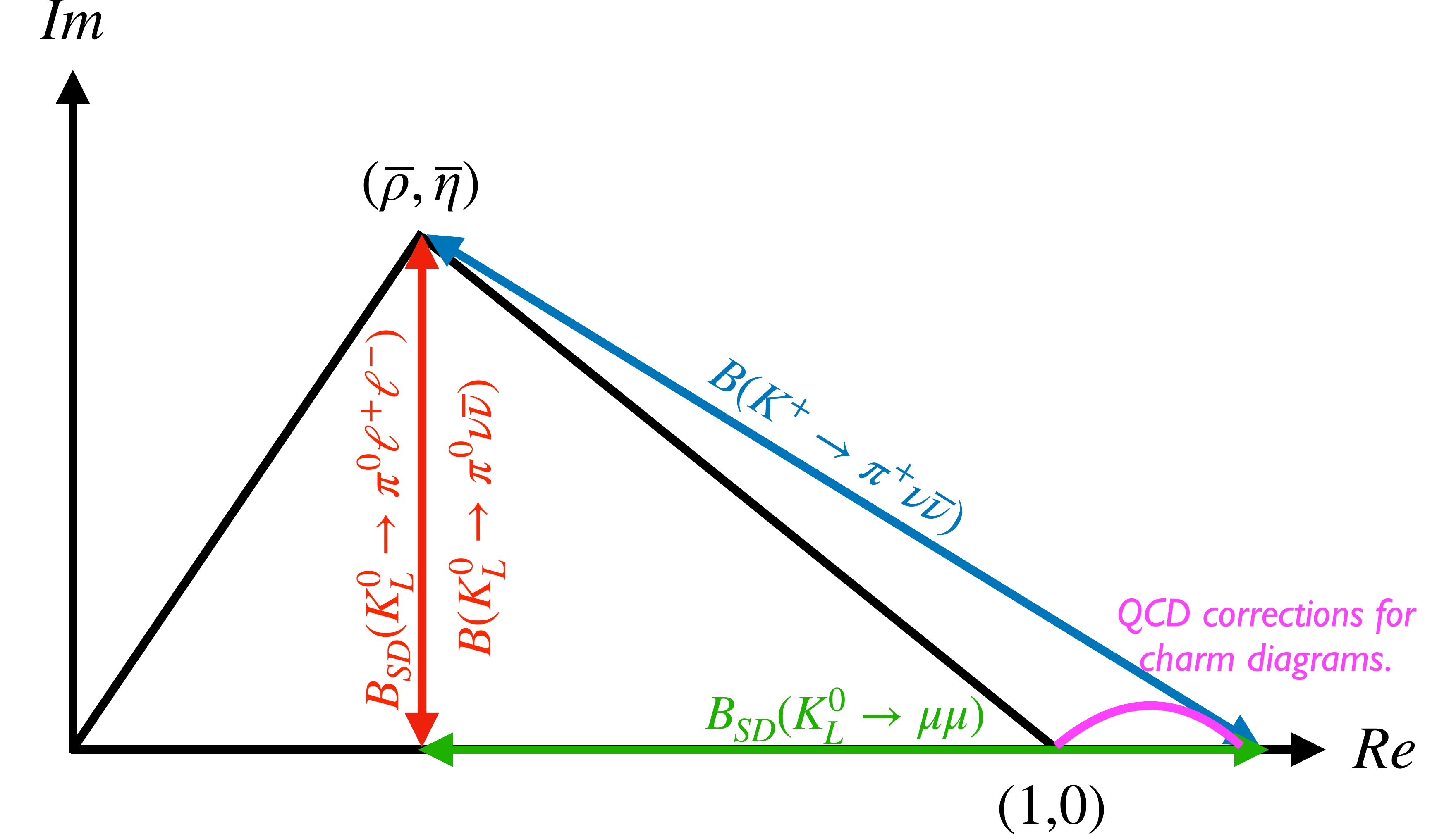}
\caption{Roles of rare kaon decays that involved in FCNC processes. SD represents the short-distance interactions.} \label{fig:rareK_role}
\end{figure}

Rare decays provide the examinations of the Standard Model (SM) if they are theoretically precise. In particular, the rare decays that involve in flavor-changing neutral current (FCNC) processes can generally have certain predictions for the New Physics (NP) contributions. Figure~\ref{fig:rareK_role} shows various rare kaon decays that have connections to the CKM parameters $\overline{\rho}$ and $\overline{\eta}$. Those measurements can shape the unitary triangle and hint NP if the CKM parameters are overconstrained. Nevertheless, both $K_L^0 \to \mu \mu$ and $K_L^0 \to \pi^0 \l^+ \l^-$ decays require a subtraction of long-distance contributions to sense the CKM parameters, and the theoretical uncertainty of long-distance interactions is known to be large. Hence, the $K \to \pi \nu \overline{\nu}$ decay, dominated by short-distance interactions, is considered to be a golden channel to search for NP due to its high theoretical precision.

The $K_L^0 \to \pi^0 \nu \overline{\nu}$ search is performed by the KOTO experiment and a branching fraction limit of $\mathcal{B}(K_L^0 \to \pi^0 \nu \overline{\nu})$ $<$ 3.0 $\times$ 10$^{-9}$ (90\% confidence level) was set \cite{KOTO_2015}. This limit is still larger than the SM prediction of (2.95 $\pm$ 0.05) $\times$ 10$^{-11}$ \cite{kpinn_SM} by two orders of magnitude. The $K^+ \to \pi^+ \nu \overline{\nu}$ search is performed by the NA62 experiment and $\mathcal{B}(K^+ \to \pi^+ \nu \overline{\nu})$ $=$ (10.6 $^{+4.0}_{-3.5}$ $|_{\text{stat.}}$) $\pm$ 0.9$_{\text{syst.}}$) $\times$ 10$^{-11}$ (68\% confidence level) was set \cite{NA62_latest}. This agrees with the Standard Model prediction of (8.60 $\pm$ 0.42 ) $\times$ 10$^{-11}$ \cite{kpinn_SM}.

\begin{figure}[h]
\centering
\includegraphics[width=80mm]{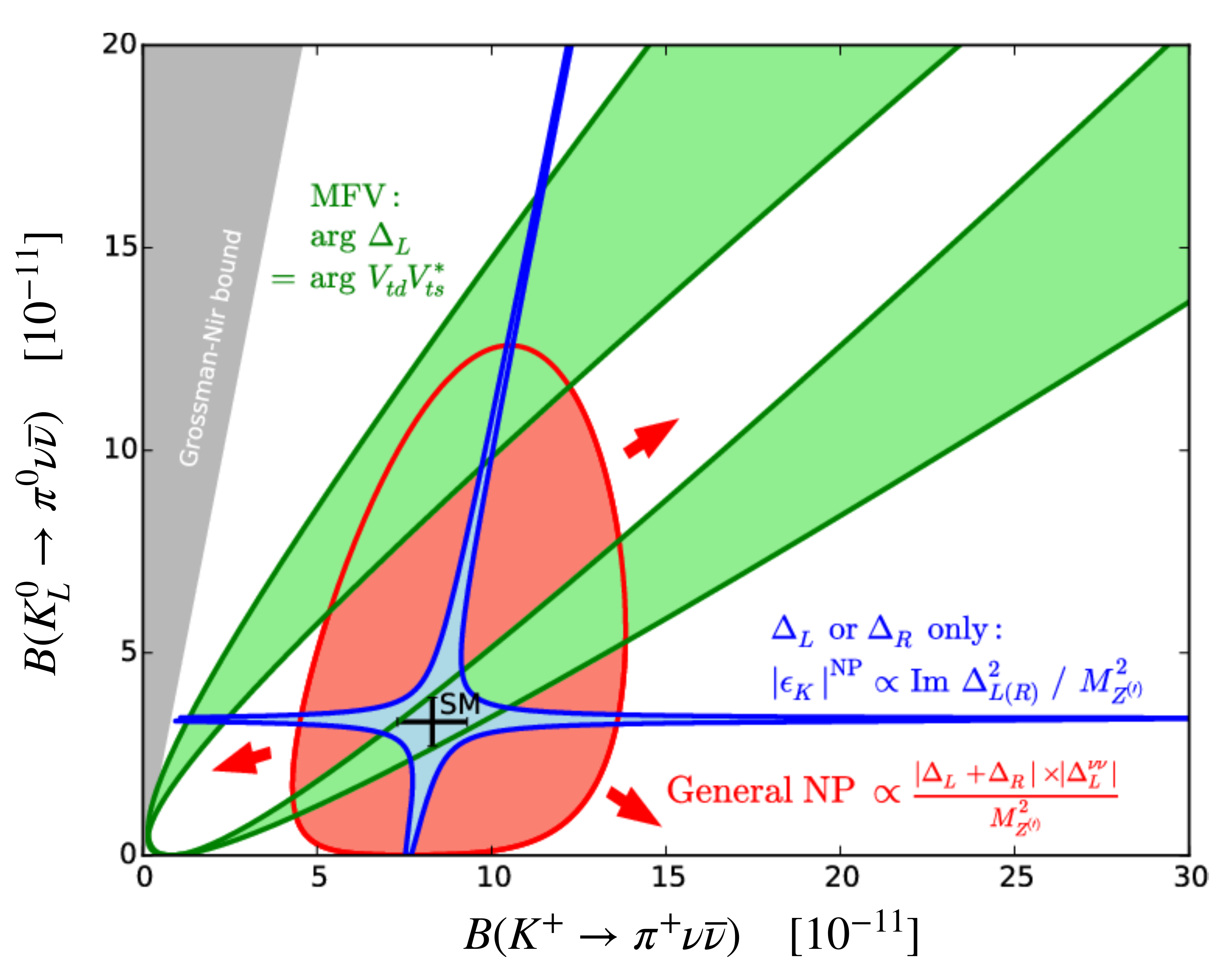}
\caption{New Physics predictions for $\mathcal{B}(K_L^0 \to \pi^0 \nu \overline{\nu})$ and $\mathcal{B}(K^+ \to \pi^+ \nu \overline{\nu})$. The grey region is ruled out by the Grossmann-Nir bound \cite{gn_bound}. The blue region indicates the models with CKM-like flavor structure. The green region indicates the models with new flavor-violating interactions where either LH or RH couplings dominate. The red region indicates the models without above constraints. (Figure courtesy of \cite{kpinn_NP}.)} \label{fig:kpinn_NP}
\end{figure}

Figure~\ref{fig:kpinn_NP} shows NP contributions to $\mathcal{B}(K_L^0 \to \pi^0 \nu \overline{\nu})$ and $\mathcal{B}(K^+ \to \pi^+ \nu \overline{\nu})$ assuming various scenarios. The Grossmann-Nir bound is a model-independent constraint governed by the weak isospin symmetry in $\Delta I = 1/2$ process \cite{gn_bound}:
\begin{align}
\mathcal{B}(K_L^0 \to \pi^0 \nu \overline{\nu}) \leq 4.3 \times \mathcal{B}(K^+ \to \pi^+ \nu \overline{\nu}). \label{eq:gn_bound}
\end{align}
By exploiting the result from NA62, the limit of $\mathcal{B}(K_L^0 \to \pi^0 \nu \overline{\nu})$ should be less than 6.3 $\times$ 10$^{-10}$ (68\% confidence level) \cite{NA62_latest}. Hence, regardless of the exceptional models that can violate the Grossmann-Nir bound, KOTO has not reached the NP sensitive region. 

The $K \to \pi \nu \overline{\nu}$ decay can also be used to search for the dark particles because the two undetected neutrinos can be interpreted as a non-SM invisible particle. The hypothesis is that dark particles may feebly interact with the SM particles. The associated scenarios include dark Higgs portal, axion portal, etc \cite{dark_theory}.  



\section{KOTO: Search for the $K_L^0 \to \pi^0 \nu \overline{\nu}$ decay}

\begin{figure*}[t]
\centering
\includegraphics[width=140mm]{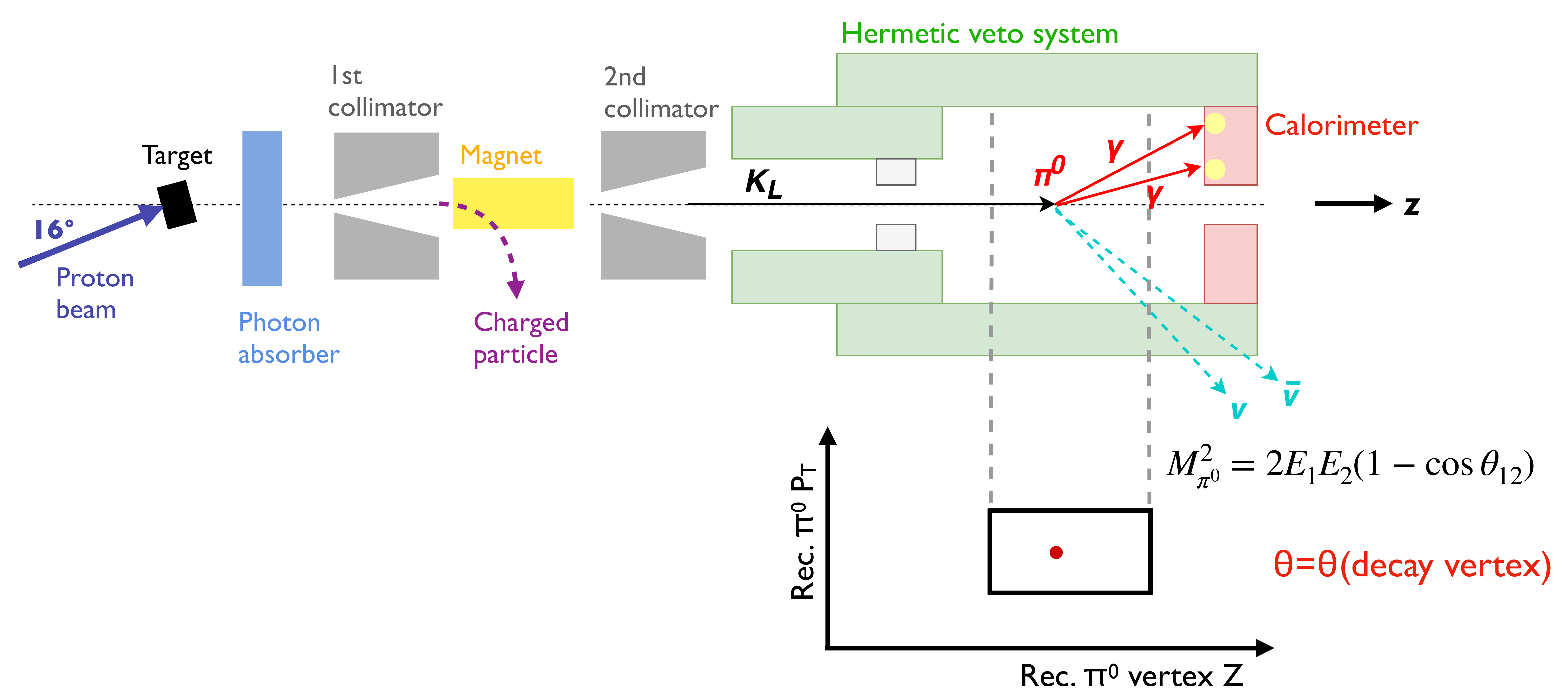}
\caption{Schematic diagram of the $K_L^0 \to \pi^0 \nu \overline{\nu}$ measurement at the KOTO experiment.} \label{fig:KOTO_schematic}
\end{figure*}

The KOTO experiment at J-PARC aims to search for the $K_L^0 \to \pi^0 \nu \overline{\nu}$ decay. The $K_L^0 \to \pi^0 \nu \overline{\nu}$ analysis principle is to measure two photons from $\pi^0$. The reconstructed $\pi^0$ transverse momentum ($P_t$) should be nonzero due to the undetected neutrinos, and nothing else should be detected. Figure~\ref{fig:KOTO_schematic} shows the schematic diagram of the $K_L^0 \to \pi^0 \nu \overline{\nu}$ measurement at KOTO. A 30-GeV proton beam collided with the target and produced a 1.4 GeV/c $K_L^0$ beam. A photon absorber and sweeping magnet were installed to eliminate photons and charged particles. Two collimators were implemented to ensure that the $K_L^0$ beam was sharp. An endcapped calorimeter was used to measure the energies and positions of the two photons from $\pi^0$. With the $\pi^0$ mass constraint, the vertex $Z$ ($Z_{vtx}$) could be obtained assuming that $\pi^0$ decays at the beam axis. The signal region was defined on $P_t$ -- $Z_{vtx}$ plane and kept blinded during the analysis until the selection criteria were fixed.

\begin{figure}[h]
\centering
\includegraphics[width=80mm]{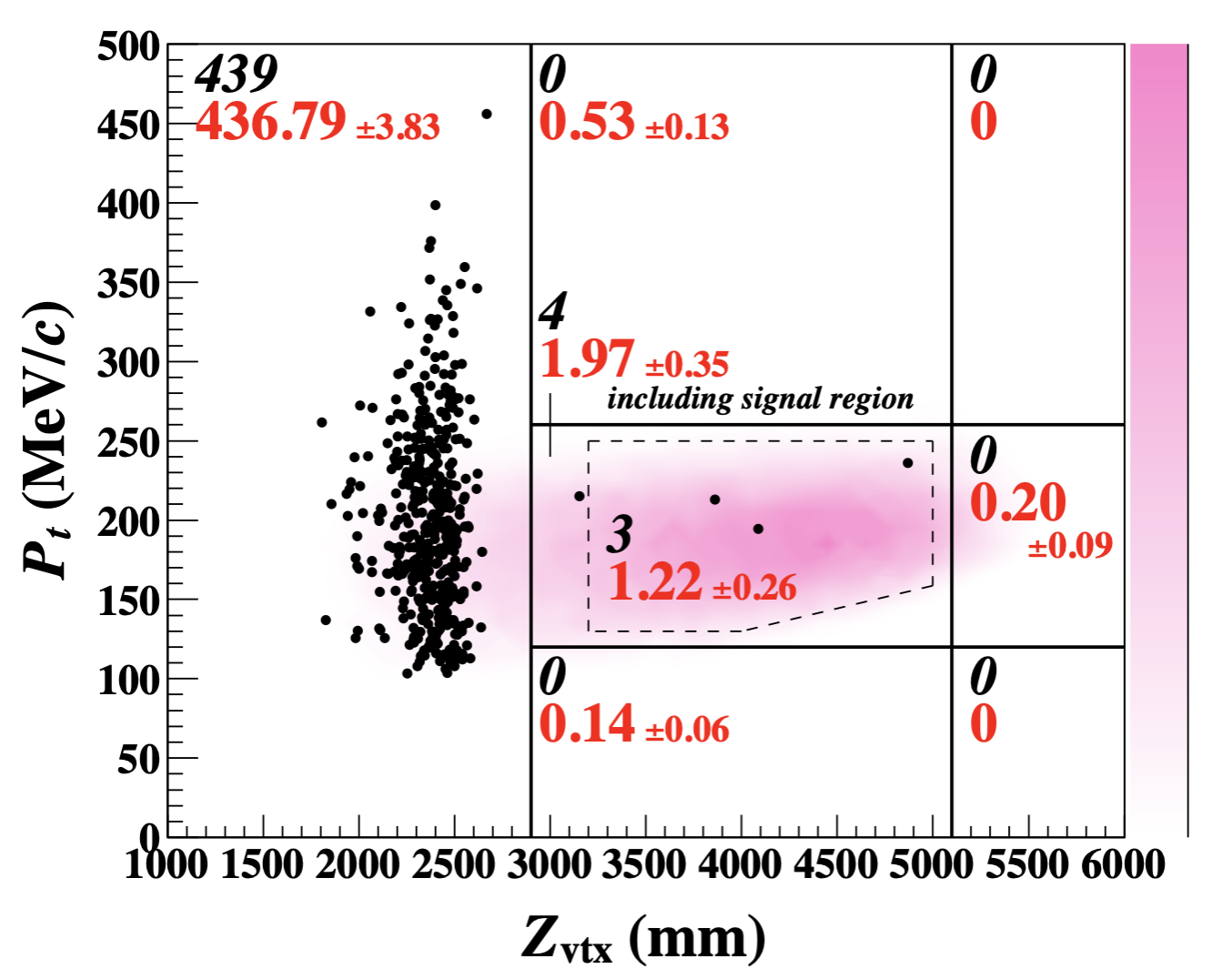}
\caption{Data distribution of $\pi^0$ $P_t$ versus $\pi^0$ vertex $Z$. Back (red) number represents the observed (predicted) number of events in the region it indicates. The magenta area represents the signal distribution.} \label{fig:KOTO_2018}
\end{figure}

KOTO achieved the single event sensitivity of 1.30 $\times$ 10$^{-9}$ using 2015 run data. With the prediction of 0.42 background events, no signal was observed. This set an upper limit of $\mathcal{B}(K_L^0 \to \pi^0 \nu \overline{\nu})$ at 90\% confidence level. KOTO further improved the single event sensitivity of 7.20 $\times$ 10$^{-10}$ using data collected from 2016 through 2018. Figure~\ref{fig:KOTO_2018} shows the result: three signal candidates were observed. Because the predicted number of background events was 1.22, the signal strength was not statistically significant. This set the upper limit of $\mathcal{B}(K_L^0 \to \pi^0 \nu \overline{\nu})$ to be 4.9 $\times$ 10$^{-9}$ at 90\% confidence level.

\begin{figure}[h]
\centering
\includegraphics[width=80mm]{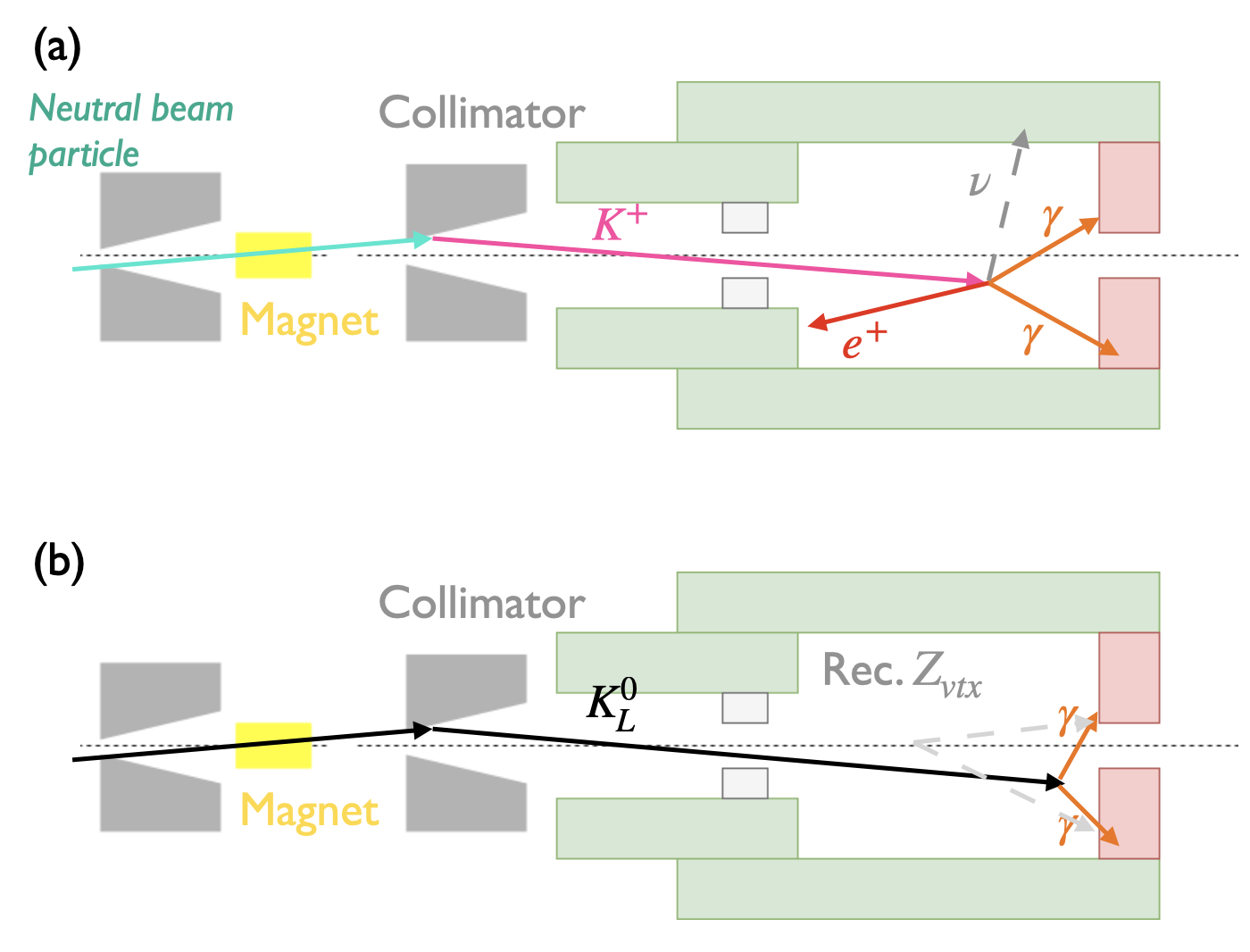}
\caption{Illustrations of KOTO major background sources. (a) A $K^+$ particle is produced at the second collimator and decays in the KOTO detector. (b) A $K_L^0$ particle is scattered at the collimator and enters the KOTO detector with a large incident angle.} \label{fig:KOTO_bg}
\end{figure}

The two major background sources found in 2016--2018 data were the $K^+$ decay and beam halo $K_L^0$ decay. Their mechanisms were illustrated in Figure~\ref{fig:KOTO_bg}. The number of background events were estimated to be 0.87 $\pm$ 0.25 and 0.26 $\pm$ 0.07 respectively. A $K^+$ particle could be produced by a beam particle at the second collimator. Because there was no magnetic field protected, a $K^+$ particle could enter the KOTO detector. Among all the $K^+$ decays, $K^+ \to \pi^0 e^+ \nu$ appeared to be the largest source. Its suppression relied on the detection of $e^+$, and a backward $e^+$ led to a large detection inefficiency. A $K_L^0$ particle could also be scattered at the collimator (halo $K_L^0$) and resulted in a large incident angle. The halo $K_L^0$ decaying into two photons could be a background if it decayed far from the beam axis and close to the calorimeter. The off-axis decay was interpreted as non-zero transverse momentum and the reconstructed vertex was shifted upstream due to the $\pi^0$ mass constraint.

KOTO plans to achieve the $K_L^0 \to \pi^0 \nu \overline{\nu}$ sensitivity by 2026. The beam intensity will be increased from 64 kW to 100 kW. In order to reduce the $K^+$ background, a plastic scintillator had been installed at the upstream of the KOTO detector in 2021. An additional magnet between the collimator and the KOTO detector will be implemented in 2023. The $K^+$ background is expected to be highly suppressed even at the SM sensitivity. For the halo $K_L^0$ background suppression, because the reconstructed incident photon angle is expected to differ from the actual one, an algorithm based on the cluster shape and incident angle is under development. A preliminary study shows an expectation of 0.4 halo $K_L^0$ background events at the SM sensitivity.
\\


\section{NA62: Search for the $K^+ \to \pi^+ \nu \overline{\nu}$ decay}

\begin{figure*}[t]
\centering
\includegraphics[width=140mm]{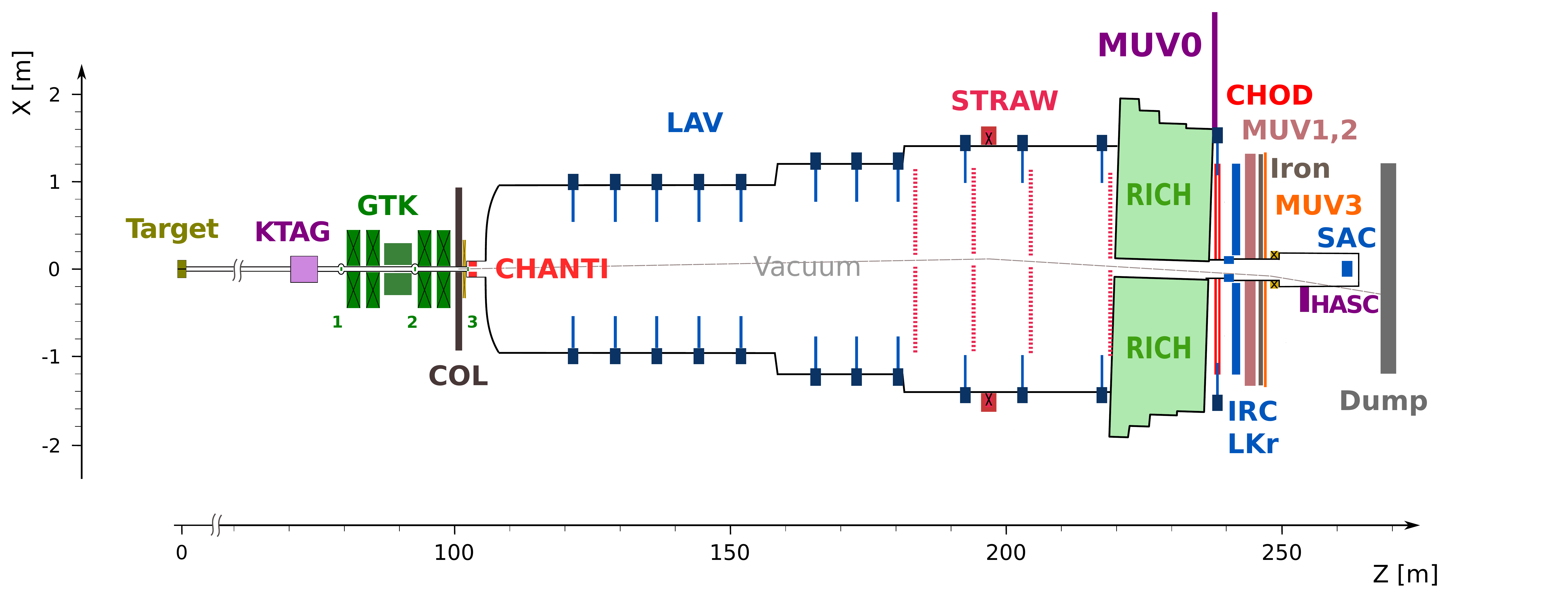}
\caption{Schematic diagram of NA62 detector (Figure courtesy of \cite{NA62_latest}).} \label{fig:NA62_detector}
\end{figure*}

The NA62 experiment at CERN aims to search for the $K^+ \to \pi^+ \nu \overline{\nu}$ decay. The $K^+ \to \pi^+ \nu \overline{\nu}$ analysis principle is to identify the incoming $K^+$ particle and the resulting $\pi^+$ particle. By measuring the $K^+$ momentum ($\overrightarrow{P_{K^+}}$) and the $\pi^+$ momentum ($\overrightarrow{P_{\pi^+}}$), the missing invariant mass ($m_{mass}$) can be calculated by
\begin{align}
   m_{miss}^2 = (\overrightarrow{P_K^+}-\overrightarrow{P_{\pi^+}})^2.
\end{align}
Because of the neutrinos, $m_{miss}$ is expected to be nonzero. Meanwhile, a $K^+ \to \pi^+ \nu \overline{\nu}$ signal requires no extra hit detected. Figure~\ref{fig:NA62_detector} shows the schematic view of the NA62 detector. A 400 GeV/c proton beam collided with the target and produced a 75 GeV/c $K^+$ beam. KTAG was a differential Cherenkov counter for $K^+$ identification. GTK was a silicon pixel beam tracker. LAV was made of lead glass for photon veto. STRAW consisted of four straw trackers with dipole magnet as a momentum spectrometer. RICH was used for the particle identification.   

\begin{figure}[h]
\centering
\includegraphics[width=80mm]{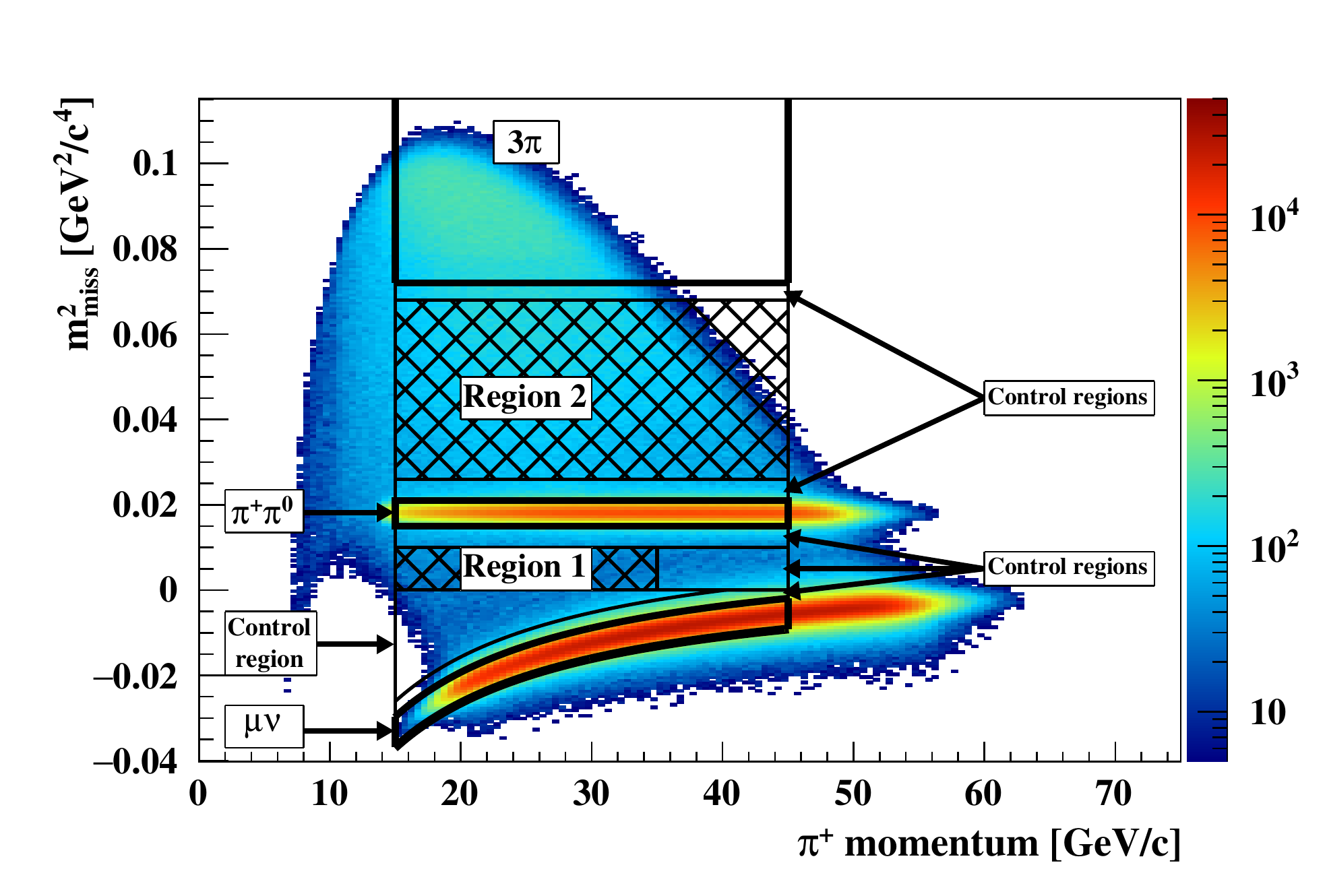}
\caption{Data distribution of $m_{miss}^2$ versus $\pi^+$ momentum. Region 1 and 2 indicate the signal region. (Figure courtesy of \cite{NA62_latest}.)} \label{fig:NA62_ana}
\end{figure}

The $K^+ \to \pi^+ \nu \overline{\nu}$ signal region is defined on the $m_{miss}^2$ -- $\overrightarrow{P_{K^+}}$ plane, as shown in Figure~\ref{fig:NA62_ana}. The high $m_{miss}^2$ region was excluded because of the $K^+ \to \pi^+ \pi^0 \pi^0$ and $K^+ \to \pi^+ \pi^+ \pi^-$ backgrounds. The two signal regions were separated by the $\pi^0$ mass band because of the $K^+ \to \pi^0 \pi^+$ background. The negative $m_{miss}^2$ region was dominated by the $K^+ \to \mu^+ \nu$ decay. The signal regions were blinded until the selection criteria were fixed.

\begin{figure}[h]
\centering
\includegraphics[width=80mm]{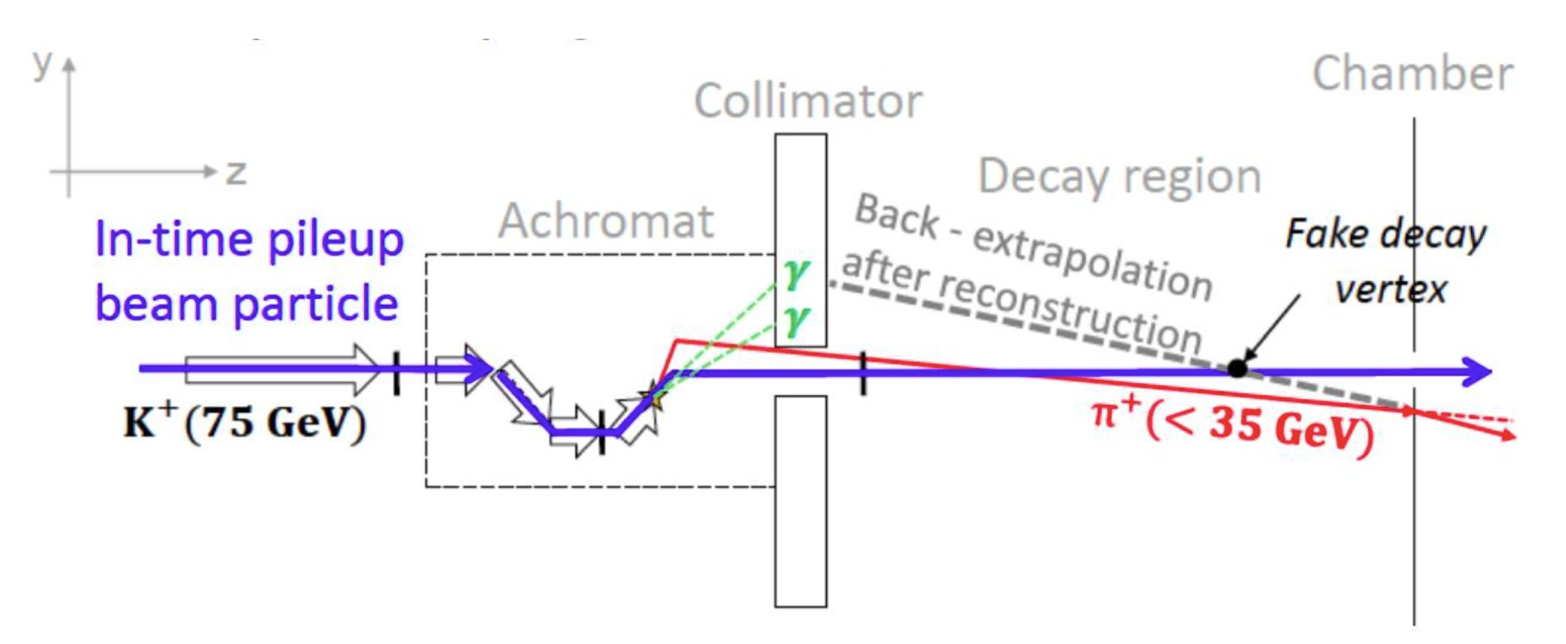}
\caption{Mechanism of the upstream background in the $K^+ \to \pi^+ \nu \overline{\nu}$ measurement.} \label{fig:KpiX_bg}
\end{figure}

The dominated background source was the upstream $\pi^+$ and its mechanism is illustrated in Figure~\ref{fig:KpiX_bg}. A $\pi^+$ particle generated upstream entered the decay region and an in-time beam particle coincided with that $\pi^+$ particle. In order to further suppress this background, the upstream collimator was upgraded in 2018 to prevent upstream $\pi^+$ from entering the decay region.    

\begin{figure}[h]
\centering
\includegraphics[width=80mm]{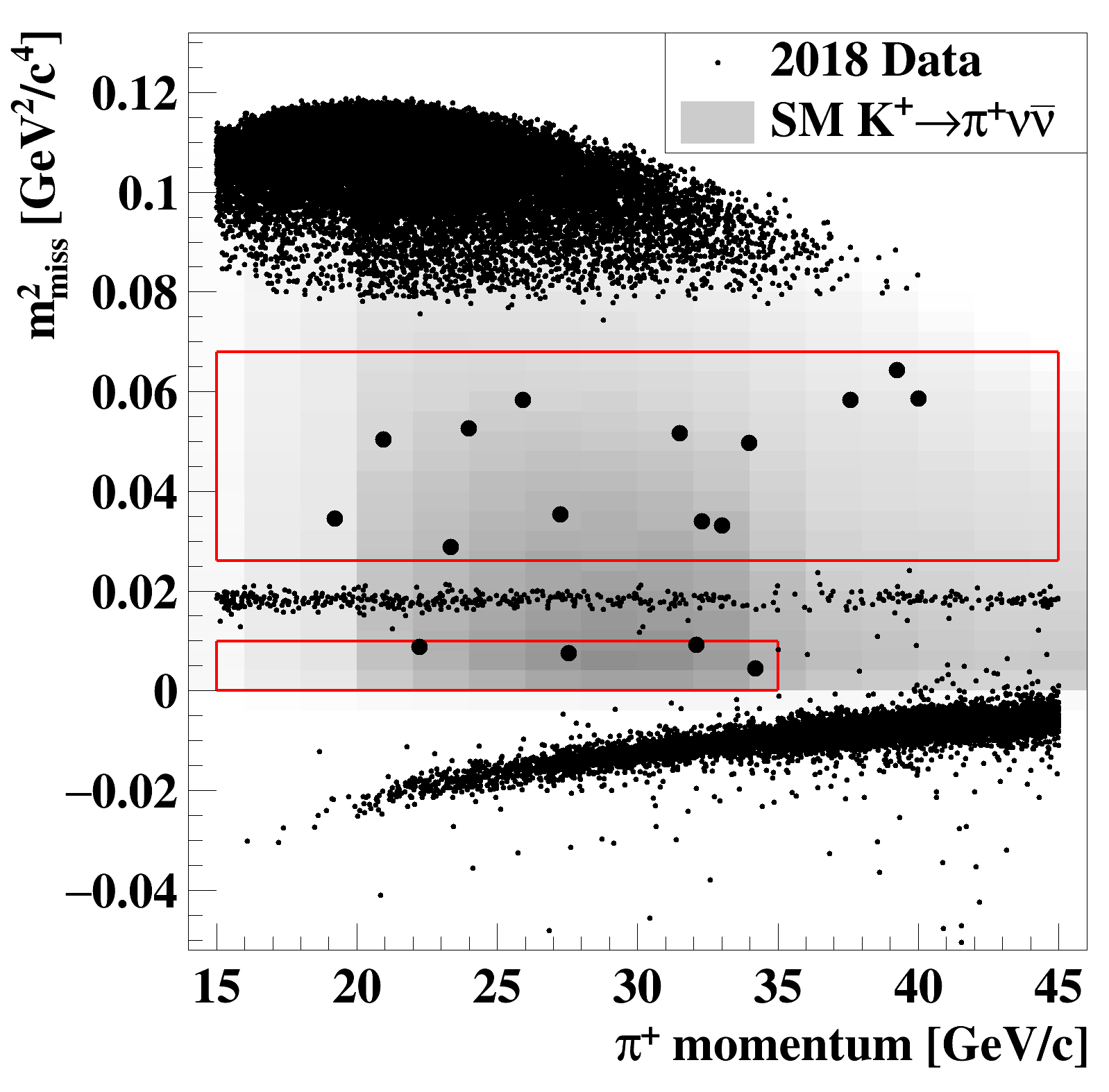}
\caption{The $K^+ \to \pi^+ \nu \overline{\nu}$ candidate events in 2018. The red boxes indicate the signal regions. (Figure courtesy of \cite{NA62_latest}.)} \label{fig:KpiX_final}
\end{figure}

Figure~\ref{fig:KpiX_final} shows the result obtained from 2018 data and 17 signals were observed. With the combination of 2016 and 2017 runs \cite{NA62_2016, NA62_2017}, 20 signals were collected and the number of background events was predicted to be 7.0. The single event sensitivity was estimated to be $(0.839 \pm 0.054) \times 10^{-11}$, which corresponds to 10.0 SM events. This set $\mathcal{B}(K^+ \to \pi^+ \nu \overline{\nu})$ to be (10.6 $^{+4.0}_{-3.5}$ $|_{\text{stat.}}$) $\pm$ 0.9$_{\text{syst.}}$) $\times$ 10$^{-11}$ at 68\% confidence level. This is equivalent with a $3.4\sigma$ observation.

Several hardware upgrades are proposed or in progress: One more beam track module (GTK) will be added and the beamline will be rearranged to swipe away upstream $\pi^+$ particles. A veto counter will be installed in the upstream region to detect upstream decay products.  An extra off-axis calorimeter (HASC) will be installed in the downstream region to further suppress the $K^+ \to \pi^0 \pi^+$ background. Furthermore, the beam intensity is anticipated to be increased. NA62 expects to measure $\mathcal{B}(K^+ \to \pi^+ \nu \overline{\nu})$ with $\mathcal{O}(10\%)$ statistical precision by the 2021--2024 runs.  


\section{Search for dark particles $X$ via $K \to \pi X$}

The $K \to \pi X$ decay can be searched using the same analysis strategy as the $K \to \pi \nu \overline{\nu}$ decay. The assumption was that $X$ lived long enough to decay outside the detector or it never decayed. 

\begin{figure}[h]
\centering
\includegraphics[width=80mm]{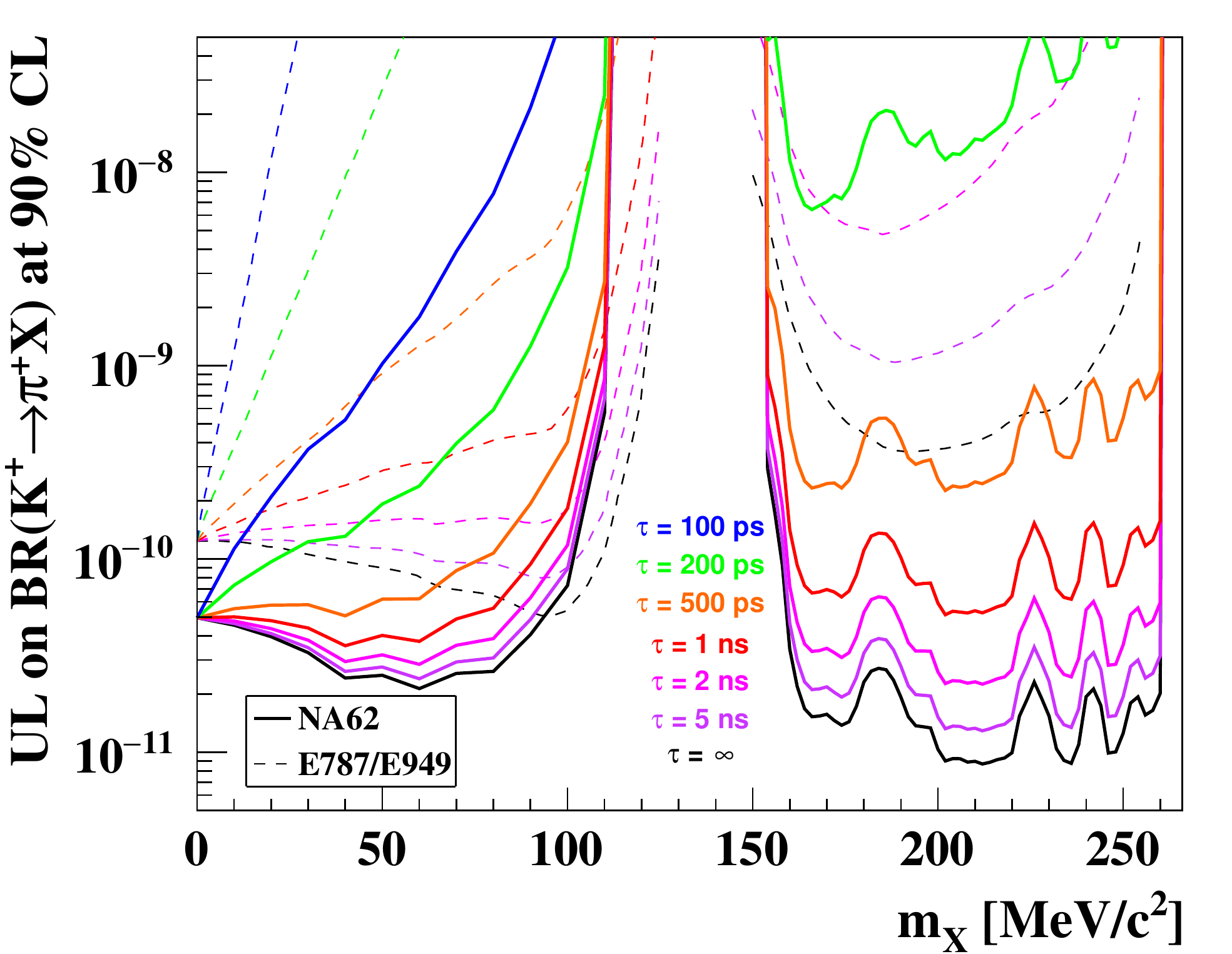}
\caption{Upper limits of $\mathcal{B}(K^+ \to \pi^+ X)$ at 90\% confidence level for different $X$ masses and lifetimes. (Figure courtesy of \cite{NA62_latest}.)} \label{fig:KpiX_UL}
\end{figure}

Figure~\ref{fig:KpiX_UL} shows the upper limits of $\mathcal{B}(K^+ \to \pi^+ \nu \overline{\nu})$ at 90\% confidence level for different $X$ masses and lifetimes. By subtracting the SM background from the observation, the branching fraction upper limits were estimated to be 3--6 $\times$ 10$^{-11}$ for the $X$ mass range of 0--110 MeV/c$^2$ and 1 $\times$ 10$^{-11}$ for the $X$ mass range of 160--260 MeV/c$^2$. According to Eq.~\ref{eq:gn_bound}, this set a stringent threshold to $\mathcal{B}(K_L^0 \to \pi^0 X)$ for the non-$\pi^0$ mass region. The $\pi^0$ mass region was excluded because of the $K^+ \to \pi^+ \pi^0$ background. Taking the detector inefficiency into account, the branching fraction upper limit of $\pi^0$ decaying into invisible particles via the $K^+ \to \pi^+ \pi^0$ decay was estimated to be 4.4 $\times$ 10$^{-9}$ at 90\% confidence level \cite{NA62_pi0invis}. 

\begin{figure}[h]
\centering
\includegraphics[width=80mm]{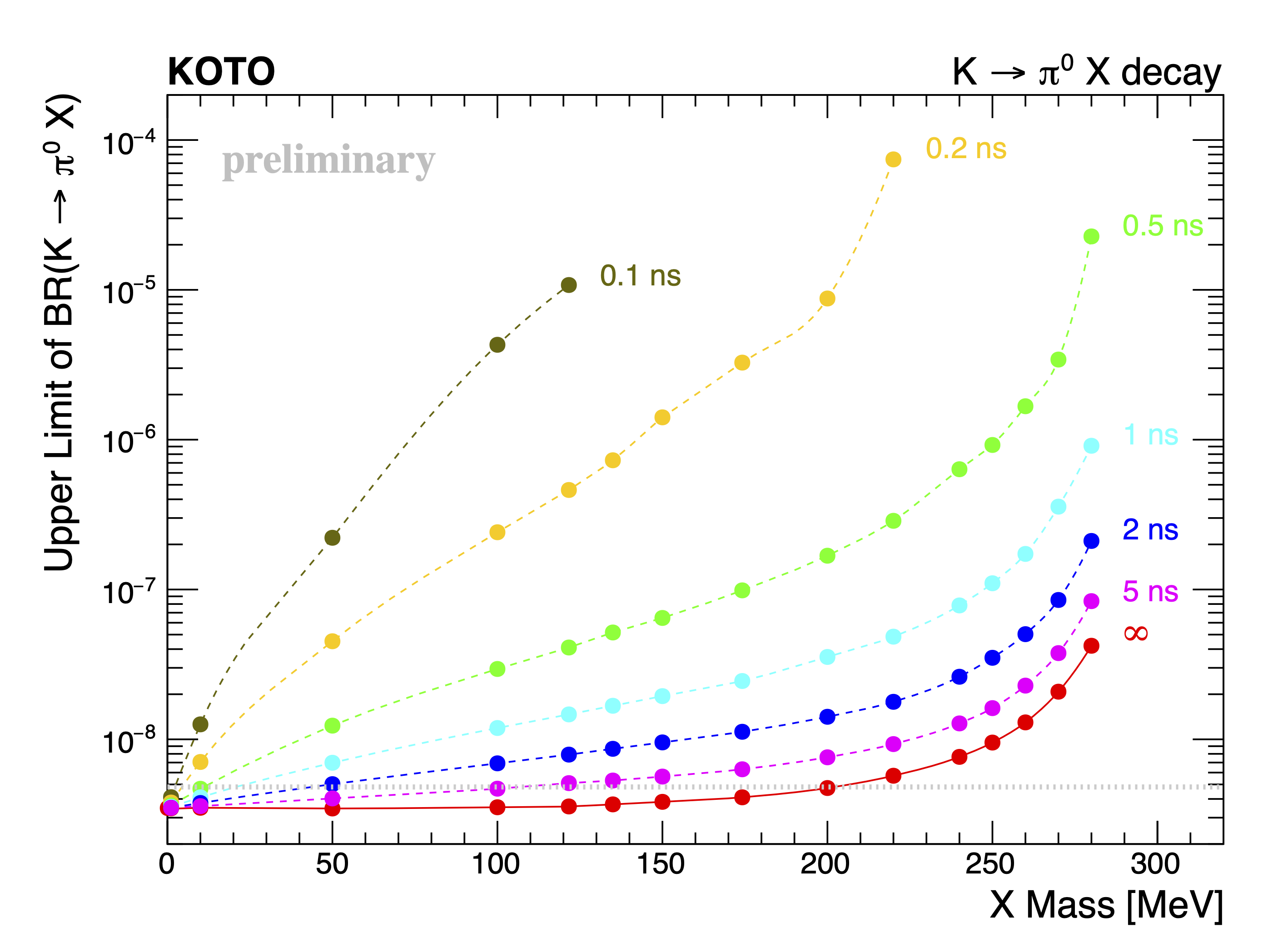}
\caption{Upper limits of $\mathcal{B}(K_L^0 \to \pi^0 X)$ at 90\% confidence level for different $X$ masses and lifetimes. The gray dashed line indicates the branching fraction limit of $K_L^0 \to \pi^0 \nu \overline{\nu}$.} \label{fig:KLpi0X_UL}
\end{figure}

Figure~\ref{fig:KLpi0X_UL} shows the upper limits of $\mathcal{B}(K_L^0 \to \pi^0 \nu \overline{\nu})$ at 90\% confidence level for different $X$ masses and lifetimes. Particularly, KOTO can examine the $\pi^0$ region in a higher sensitivity. If $X$ has mass of $\pi^0$, the branching fraction upper limit of $K_L^0 \to \pi^0 X$ was estimated to be 3.7 $\times$ 10$^{-9}$ at 90\% confidence level (preliminary). 

\section{Future kaon experiments}

Two experiments, KOTO step-2 \cite{koto2} and KLEVER \cite{klever}, are proposed to measure $\mathcal{B}(K_L^0 \to \pi^0 \nu \overline{\nu})$ in a higher precision. The goal is to achieve the single event sensitivity of $\mathcal{O}(10^{-13})$. 

KOTO step-2, as an upgraded version of current KOTO, is expected to have 2.4 times higher $K_L^0$ flux because the angle between the primary proton beam and the $K_L^0$ beam decreases from 16$^{\circ}$ to 5$^{\circ}$. The resulting $K_L^0$ momentum is expected to peak at 3 GeV/c and thus a larger detector is required to keep the decay volume fully enclosed by veto counters. With this design, 35 SM events with 56 background events are expected, and it corresponds to a branching fraction precision of 27\%. In the earliest scenario, KOTO step-2 will start from 2029.

KLEVER at CERN SPS is designed to collect 60 SM events with the signal-to-background ratio of 1, and it correponds to a branching fraction precision of 20\%. The 40-GeV $K_L^0$ beam will be produced by a 400-GeV proton beam. The detector consists of an electromagnetic calorimeter used to reconstruct $\pi^0 \to 2\gamma$ and layers of rings along the decay region to veto photons. KLEVER aims to start data taking from 2026 (after long shut down in LHC).


\section{Conclusion}

The rare kaon decay $K \to \pi \nu \overline{\nu}$ is the golden channel for the New Physics search due to its high theoretical precision. KOTO is searching for the $K_L^0 \to \pi^0 \nu \overline{\nu}$ decay and achieved a limit of  $\mathcal{B}(K_L^0 \to \pi^0 \nu \overline{\nu})$ $<$ 3.0 $\times$ 10$^{-9}$ (90\% confidence level) using 2015 data. The later data set collected from 2016 through 2018 revealed that the $K^+$ and halo $K_L^0$ backgrounds were profound. A hardware upgrade and new algorithms have been developed to suppress the two backgrounds to an acceptable level. KOTO expects to achieve the SM sensitivity at 2026. NA62 is searching for the $K^+ \to \pi^+ \nu \overline{\nu}$ decay and showed $\mathcal{B}(K^+ \to \pi^+ \nu \overline{\nu})$ $=$ (10.6 $^{+4.0}_{-3.5}$ $|_{\text{stat.}}$) $\pm$ 0.9$_{\text{syst.}}$) $\times$ 10$^{-11}$ at 68\% confidence level. This measurement agrees with the SM prediction. The detector is upgraded to reduce the background further. NA62 expects to achieve $\mathcal{O}(10\%)$ accuracy of $\mathcal{B}(K^+ \to \pi^+ \nu \overline{\nu})$ using 2021--2014 run data. The dark particle $X$ search via $K \to \pi X$ was also performed by KOTO and NA62. The upper limit of $\mathcal{B}(K^+ \to \pi^+ X)$ was set to be 3--6 $\times$ 10$^{-11}$ for the $X$ mass range of 0--110 MeV/c$^2$ and 1 $\times$ 10$^{-11}$ for the $X$ mass range of 160--260 MeV/c$^2$. KOTO complemented the examination in the $\pi^0$ mass region; the upper limit of $\mathcal{B}(K_L^0 \to \pi^0 X)$, where $X$ has mass of $\pi^0$, was set to be 3.7 $\times$ 10$^{-9}$ at 90\% confidence level (preliminary). In the future, KOTO step-2 and KLEVER are proposed to measure $\mathcal{B}(K_L^0 \to \pi^0 \nu \overline{\nu})$ at the single event sensitivity of $\mathcal{O}(10^{-13})$. 



\bigskip 
\bibliography{ref}

\end{document}